\documentclass{article}
\usepackage[utf8]{inputenc}

\usepackage{graphicx,setspace}
\graphicspath{ {images/} }
\usepackage{caption}
\usepackage{subcaption}
\usepackage[a4paper,width=150mm,top=23mm,bottom=23mm]{geometry}
\usepackage{mathtools}
\usepackage{enumerate}
\usepackage{listings}
\usepackage{color}
\usepackage{esint}
\usepackage{soul}
\usepackage[table,xcdraw]{xcolor}
\usepackage{pifont}
\usepackage{float}
\usepackage{url}
\usepackage{varwidth}
\usepackage{dutchcal}
\usepackage{mathrsfs}
\usepackage{parskip}
\usepackage{amsmath,amsthm,amssymb}

\numberwithin{equation}{section}

\usepackage{pgf, tikz}
\usetikzlibrary{arrows, automata}
\usepackage{framed}

\allowdisplaybreaks
\usepackage{bbm}
\usepackage[makeroom]{cancel}

\usepackage[maxbibnames=99,style=numeric,backend=biber,sorting=nyt, bibstyle=authoryear]{biblatex}

\DeclareFieldFormat
  [article,book,incollection,inproceedings,misc,thesis,unpublished]
  {title}{#1\isdot}

\AtNextBibliography{\small}
\DeclareNameAlias{sortname}{family-given}
\DeclareBibliographyCategory{}

\DeclareFieldFormat{labelnumberwidth}{\mkbibbrackets{#1}}
\defbibenvironment{bibliography}
  {\list
     {\printtext[labelnumberwidth]{%
    \printfield{prefixnumber}%
    \printfield{labelnumber}}}
     {\setlength{\labelwidth}{\labelnumberwidth}%
      \setlength{\leftmargin}{\labelwidth}%
      \setlength{\labelsep}{\biblabelsep}%
      \addtolength{\leftmargin}{\labelsep}%
      \setlength{\itemsep}{\bibitemsep}%
      \setlength{\parsep}{\bibparsep}}%
      }
  {\endlist}
  {\item}

\renewbibmacro{in:}{%
  \ifentrytype{article}
    {}
    {\bibstring{in}%
     \printunit{\intitlepunct}}}
     
\usepackage{xpatch}
\xpatchbibmacro{cite}{\printnames{labelname}}%
{\ifciteseen{\printnames{labelname}}{\printnames[][-\value{listtotal}]{labelname}}}
{}{}

\xpatchbibmacro{textcite}{\printnames{labelname}}%
{\ifciteseen{\printnames{labelname}}{\printnames[][-\value{listtotal}]{labelname}}}
{}{}

\newcommand{\IP}{\mathbb{P}}

\newcommand{\IE}{\mathbbm{E}}
\newcommand{\E}{\mathbbm{E}}

\renewcommand{\leq}{\leqslant}
\renewcommand{\geq}{\geqslant}

\newcommand\numberthis{\addtocounter{equation}{1}\tag{\theequation}}
\makeatletter
\newcommand*{\rom}[1]{\expandafter\@slowromancap\romannumeral #1@}
\makeatother

\def\given{\typeout{Command 'given' should only be used within bracket command}}
\newcounter{@bracketlevel}
\def\@bracketfactory#1#2#3#4#5#6{
\expandafter\def\csname#1\endcsname##1{%
\addtocounter{@bracketlevel}{1}%
\global\expandafter\let\csname @middummy\alph{@bracketlevel}\endcsname\given%
\global\def\given{\mskip#5\csname#4\endcsname\vert\mskip#6}\csname#4l\endcsname#2##1\csname#4r\endcsname#3%
\global\expandafter\let\expandafter\given\csname @middummy\alph{@bracketlevel}\endcsname
\addtocounter{@bracketlevel}{-1}}%
}
\def\bracketfactory#1#2#3{%
\@bracketfactory{#1}{#2}{#3}{relax}{1mu plus 0.25mu minus 0.25mu}{0.6mu plus 0.15mu minus 0.15mu}
\@bracketfactory{b#1}{#2}{#3}{big}{1mu plus 0.25mu minus 0.25mu}{0.6mu plus 0.15mu minus 0.15mu}
\@bracketfactory{bb#1}{#2}{#3}{Big}{2.4mu plus 0.8mu minus 0.8mu}{1.8mu plus 0.6mu minus 0.6mu}
\@bracketfactory{bbb#1}{#2}{#3}{bigg}{3.2mu plus 1mu minus 1mu}{2.4mu plus 0.75mu minus 0.75mu}
\@bracketfactory{bbbb#1}{#2}{#3}{Bigg}{4mu plus 1mu minus 1mu}{3mu plus 0.75mu minus 0.75mu}
}
\bracketfactory{clc}{\lbrace}{\rbrace}
\bracketfactory{clr}{(}{)}
\bracketfactory{cls}{[}{]}
\bracketfactory{abs}{\lvert}{\rvert}
\bracketfactory{norm}{\Vert}{\Vert}
\bracketfactory{floor}{\lfloor}{\rfloor}
\bracketfactory{ceil}{\lceil}{\rceil}
\bracketfactory{angle}{\langle}{\rangle}

\theoremstyle{plain}

\theoremstyle{definition}

\usepackage{enumitem}

\usepackage{titlesec}
\titleformat{\section}
  {\centering}{\thesection}{1em}{\MakeUppercase}

\usepackage[breaklinks=true]{hyperref}
\hypersetup{
colorlinks=true,
linkcolor=blue,
urlcolor=blue,
citecolor=black
}

\title{Properties of the `friend of a friend' model for network generation}
\author{Tiffany Y.\ Y.\ Lo\thanks{Department of Mathematics, Uppsala University. Email address: \url{yin_yuan.lo@math.uu.se}}, Watson Levens\thanks{Department of Mathematics, University of Dar es Salaam. Email address: \url{watsonlevens@udsm.ac.tz}}, David J.\ T.\ Sumpter\thanks{Department of Information Technology, Division of Systems and Control, Uppsala University. Email address: \url{david.sumpter@it.uu.se}}}
\date{\today}

\begin{document}

\maketitle

\begin{abstract}
The way in which a social network is generated, in terms of how individuals attach to each other, determines the properties of the resulting network. Here we study an intuitively appealing `friend of a friend' model, where a network is formed by each newly added individual attaching first to a randomly chosen target and then to $n_q\geq 1$ randomly chosen friends of the target, each with probability $0<q\leq1$. We revisit the master equation of the expected degree distribution for this model, providing an exact solution for the case when $n_q$ allows for attachment to all of the chosen target's friends (a case previously studied by \cite{lambiotte2016}), and demonstrating why such a solution is hard to obtain when $n_q$ is fixed (a case previously studied by \cite{Levens2022}.) In the case where attachment to all friends is allowed, we also show that when $q<q^*\approx0.5671$, the expected degree distribution of the model is stationary as the network size tends to infinity. We go on to look at the clustering behaviour and the triangle count, focusing on the cases where $n_q$ is fixed. 

\end{abstract}

\section{Introduction}

Social network models attempt to capture interpersonal connections, both in the digital platforms and the physical world.  There are wide range of generative models which describe how interaction between individual components 
lead to the generation of certain forms of network. The most well-known example is perhaps the preferential attachment mechanism popularised by Barab\'asi-Albert \cite{barabasi1999emergence},  where individuals rich in connections are more likely to gain new connections. Another popular example is the Watts-Strogatz model \cite{watts1998collective}, where the network is generated in such a way that through edge rewirings, individuals are allowed to form connections with other individuals that do not belong to the same `local clusters' as themselves. 


Although the preferential attachment model produces the power-law degree distribution and the Watts-Strogatz model exhibits the `small-world' properties, both of which are key features of social networks, the growth mechanism of neither models are based on local interactions. That is, these models are generated using information on the {\it whole} network. For example, when a new individual connects to one or multiple existing individuals in a standard  preferential attachment network, the probability that the new individual connects to any individual is proportional to the current degree of the existing individual. In contrast, interactions between individuals in social networks depend only on local information. Copying models are one of several generative network models that remedy the limitation of non-local interactions \cite{bianconi2014triadic, chung2003duplication, kleinberg1999web, krapivsky2005network, lambiotte2016, vazquez2000knowing}. Some common assumptions of the copying models include that individuals have limited information about the network so individuals make connection decisions based on their local view of the network, and that new individuals 
have a tendency to connect to individuals with higher degrees (the so called `friendship paradox'.) 



One example of such models for explaining the emergence of social networks is introduced by \textcite{lambiotte2016}. In this model, a new individual $B$ is introduced to the network at each discrete time step and connects to an existing individual $A$. Then, $B$ is introduced to all the friends of $A$ in the network, and $B$ becomes friends with each of them with probability $0\leq q\leq 1$, with each decision made independently. When $q=0$, the model reduces to the well-known random recursive tree; and when $q=1$, the model yields a complete graph if the initial network is a complete graph. For $0<q\leq 1$, the authors of \cite{lambiotte2016} show that the average number of edges in a network with $t$ edges exhibits the following phase transition. When $q<1/2$, it is of order $t$; when $q=1/2$, it is $t\ln t$; and when $q>1/2$, it is $t^{2p}$. This indicates that the resulting network is dense when $q>1/2$, and is sparse otherwise. They also show that the expected numbers of cliques in the network have a similar behaviour, with phase transitions observed at several other values of $q$. 

Generative models similar to that of \cite{lambiotte2016} have been introduced in \textcite{chung2003duplication} and \textcite{vazquez2003modeling} for explaining protein interaction networks, where the only difference is that the new node never connects to the randomly chosen target node. It has been proved in \cite{BarbourL2022,bebek2006degree, hermann2016,jordan2018} that when $q\leq q^*\approx 0.5671$, almost all nodes in the network are isolated, indicating that the model is not suitable for fitting social and protein interaction networks for small values of $q$. On the other hand, there can be a non-trivial proportion of nodes whose degrees are increasing with the network size when $q>q^*$. However, a further analysis can be conducted by considering only the connected component of the network, as done in \cite{ BarbourL2022,jacquet2020, jordan2018}. In particular, the degree distribution of a randomly chosen node in the network, conditional on it being part of the connected component, is proved in \cite{jacquet2020} to display a power-law behaviour when the network size is large and $0<q<e^{-1}$, where the exponent $\beta(q)>2$ satisfies $q^{\beta-2}+\beta-3=0$. A modification of the model has also been proposed by \cite{bebek2006degree,pastor2003evolving}, where the newly added node may attach to individuals that are neither the randomly chosen target nor its friends, in addition to copying the links of the target node. As shown in \cite{bebek2006degree, BarbourL2022}, for a suitable range of $q$, the expected degree distribution of the modified network has a limit that is not a point mass at zero, as the network size tends to infinity.


 
The `friend of a friend' model considered in \textcite{Levens2022} also describes the formation of social networks using a series of local individual interactions and builds on the work of Dorogovtsev \cite{dorogovtsev2002evolution}, Jackson and Rogers \cite{jackson2007meeting} and Krapivsky and Redner \cite{krapivsky2001organization,krapivsky2005network,krapivsky2017emergent}. In this model, the initial network has three connected individuals $A$, $B$ and $C$. A person $D$, who joins a social network initiates his/her friend selection by choosing a person, say $B$, randomly from the network without considering $B$'s social connections. Person $B$ then selects another individual randomly among his/her friends, say person C, as a potential friend for $D$. Then, with probability $p$ (respectively $q$), person $D$ becomes friends with person $B$ (respectively person $C$), where in particular, $D$ befriends both individuals with probability $p q$. 
It is shown in \cite{Levens2022} that the model generates networks that exhibit power-law degree distributions with empirical exponents ranging from 1.5 upwards and small-world clustering, mirroring those observed in real-world networks. When $p$ is small enough and $q=1$, the model also produces networks characterised by the presence of individuals with an exceptionally high degree of connectivity within the network and such individuals are also commonly observed in social networks. Indeed, the model is a special case of the model studied in \textcite{saramaki2004scale} when $p$ equals $0$ and $q$ equals $1$, in which case the proportion of leaves becomes dominant as the network expands; see also \cite{cannings2013random} for more mathematical details behind this phenomenon.


The key difference in the generative mechanism of the model of \cite{lambiotte2016} and the `friend of a friend' model is in how many friends of the target individual are chosen as friends of the newly added individual. Indeed, we can construct a model that includes these two models as its special case. Let $0<q\leq 1$ and let $n_q\geq 1$ be a positive integer. In this more general model, when $n_q$ is fixed, the initial network is a complete graph with $(n_q+2)$ individuals; and when $n_q$ is equal to the degree of the randomly chosen target\footnote{For convenience, hereafter we refer to this case by simply writing $n_q=Deg^R(t)$, with $R=R(t)$ being the randomly chosen vertex at time $t$ and $Deg^R(t)$ being the degree of individual $R$ when the network has $t$ individuals.}, it is a triangle (this chosen purely to simplify our arguments later.) The individuals are labelled $\{1,2,3,\dots\}$ and are sequentially introduced over time. At each time step $t\geq n_q+3$ (or $t\geq 4$ when $n_q=Deg^R(t)$), the newly added individual, labelled $t$, attaches to a target individual, say $u$, that is randomly chosen from the existing network of $t-1$ individuals. Individual $t$ further chooses an $n_q$-subset of the friends of $u$, and connects to each of them with probability $q$, and the decisions are made independently. When $n_q=1$, the model corresponds to the `friend of a friend' model with $p=1$, where each individual in the network has at least one friend. When $n_q=Deg^R(t)$, it is easy to see that the model coincides with that of \cite{lambiotte2016}, where all friends of the target may form a connection with the new individual. In the case of $n_q\geq 2$ however, the randomly chosen target at the subsequent steps may not have enough friends to be sampled from when $q<1$, since the previously added individuals may or may not attach to the chosen friends of their targets. But when $q=1$, this issue is remedied by initialising the network with a complete graph on $(n_q+2)$ individuals, so that each new individual in the network has at least $n_q+1$ friends to choose from. 




For convenience, we will simply refer to the model above as the `friend of a friend' model and  investigate the degree distribution, where we separately consider the cases where $n_q$ is fixed and where $n_q=Deg^R(t)$. When $n_q\geq 2$, we assume further that $q=1$. We also investigate the triangle count and the clustering behaviour when $n_q$ is fixed. As social networks are characterised by power-law degree distribution and small-world clustering, these properties of the model are of particular importance. On the other hand, counts of small graphs can be used as summary statistics and over-represented small graphs (the so-called motifs) have been conjectured as the building blocks of real-world networks; see e.g.\ \cite{ali2014alignment,liu2011evidence}. Here we focus on the triangle count, since its analysis is also closely related to that of the clustering behaviour. 

The remainder of this article is organised as follows. In Section \ref{se:degree}, we derive the master equation for the expected degree distribution. Using the formula, we will demonstrate that when $n_q$ is fixed, the analysis of the recursion equation is in general, difficult, mostly due to the fact that the recursion also involves on the degrees of the friends of the random target individuals. In this case, we will rely on means of approximations to solve the recursion, and compare the result to simulations. When $n_q=Deg^R(t)$, the recursion can be analysed without using any approximation. In particular, we show that the expected degree distribution is stationary only  when $q<q^*$, where $q^*\approx 0.5671$. In~Section \ref{se:tri}, we study the triangle count, showing that when $n_q\geq 1$, its expected number increases linearly with the number of individuals in the network. This is in contrast to the behaviour of the triangle count in the case of $n_q=Deg^R(t)$, whose growth with respect to the network size is sensitive to the choice of $q$, as already shown in \cite{lambiotte2016}. When $n_q=1$, we further show that the triangle count obeys the law of large numbers and is asymptotically normal. In Section~\ref{se:cluster}, we investigate the clustering coefficient. The analysis of the clustering coefficient is highly non-trivial, and we show this by deriving a suitable master equation when $n_q$ is fixed. However, when $n_q=1$ and $q=1$, it is possible to obtain an exact expression for the clustering coefficient. When $n_q\geq 2$, we analyse the clustering of the randomly chosen target individual by means of approximation. 
We conclude with a short discussion on our results and possible future directions in Section \ref{se:discussion}.



\section{Methods and Results}\label{se:res}

\subsection{The expected degree distribution}\label{se:degree}
\subsubsection{Deriving the master equation} 

We now derive a master equation for the \emph{expected} degree distribution of the network. Let $G_t$ be the `friend of friend' network right after adding the $t$-th individual and his/her connections to individuals who have joined the network before him/her. Let $N_k(t)$ be the number of individuals of degree $k$ in $G_t$. In the case where $n_q\geq 2$, we emphasise that we further set $q=1$, often times without explicitly saying so when we analyse the more general case $n_q\geq 1$. 

\subsubsection*{Changes in degree of existing individuals}

When adding the new individual $t+1$ to the existing network with $t$ individuals, an individual in the network with degree $k-1$ can increase his/her degree to $k$ in two ways. Firstly, the individual could be uniformly chosen as the friend of the newly added individual, which happens with probability
    \begin{equation}
        \frac{N_{k-1}(t)}{t}.
    \end{equation}

The second way of increasing degree is that a friend of an individual $u$ of degree $k-1$ in the network befriends the newly introduced individual $t+1$. We write $S_{k}(t)$ to denote the set of individuals that have $k$ friends at time $t$, and $F_{u}(t)$ to denote the set of friends of individual $u$ at time $t$. Then, given the network at time $t$, the expected change in the count is
    \begin{align*}
    & \sum_{u\in S_{k-1}(t)}\IP[\text{a friend of individual $u$ is chosen and the edge $u\sim t$ is created}\mid G_t]\\
    &\quad= \sum_{u\in S_{k-1}(t)}\sum_{v\in F_{u}(t)}\IP[\text{friend $v$ of individual $u$ is chosen and the edge $u\sim t$ is created}\mid G_t]\\
    &\quad=  \sum_{u\in S_{k-1}(t)}\sum_{v\in F_{u}(t)}\frac{q}{t}\frac{\binom{Deg_{v}(t)-1}{n_q-1}}{ \binom{Deg_{v}(t)}{n_q}}. \numberthis \label{eq:FoF1}
    \end{align*}
To understand the last probabilities inside the double sum above, consider that given that $v$ is chosen as the target individual (which happens with probability $1/t$), the probability that friend $u$ of $v$ is further chosen as a potential new friend is $\binom{Deg_{v}(t)-1}{n_q-1}$ divided by $\binom{Deg_{v}(t)}{n_q}$. The probability that the potential friend is attached to is then $q$.

At this point, whether individual $t+1$ also befriends $u$ depends on the parameters $q$ and $n_q$. In the case where $n_q = Deg^R(t)$, all friends of $u$ are potential friends for $t$ and we can simplify, such that,
    \begin{align*}
    &\sum_{u\in S_{k-1}(t)}\IP[\text{a friend of individual $u$ is chosen and the edge $u\sim t$ is created}\mid G_t]\\
    &\quad=\sum_{u\in S_{k-1}(t)}\sum_{v\in F_{u}(t)} \frac{q}{t}, \\
    &\quad=\sum_{u\in S_{k-1}(t)} \frac{(k-1)q}{t} \quad\text{since every individual $u$ has $(k-1)$ friends,}\\
    &\quad=\frac{q(k-1)N_{k-1}(t)}{t}, \quad\text{the size of $S_{k-1}(t)$ is exactly $N_{k-1}(t)$}.\numberthis\label{eq:secondary}
    \end{align*}
In the case where $n_q$ is fixed, we have:
\begin{align}
&\sum_{u\in S_{k-1}(t)}\IP[\text{a friend of individual $u$ is chosen and the edge $u\sim t$ is created}\mid G_t] \nonumber \\
&\quad \approx \sum_{u\in S_{k-1}(t)}\sum_{v\in F_{u}(t)}\frac{1}{t}\frac{qn_q}{Deg_{v}(t)}  .\label{eq:generalnq}
\end{align}
and no immediate simplification is possible, because $Deg_{v}(t)$ also depends on $v$. To do this, one may derive another set of master equations for $\E[Deg_{u}(t)]$ and approximate $Deg_{u}(t)$ with $\E[Deg_{u}(t)]$. 

An alternative, taken by \textcite{Levens2022}, is to assume that $\E[Deg_{u}(t)]$ is approximated by the expected average degree at time $t$. Since we add an average of $1+q n_q$ edges to the graph at each time step, the average degree of the system is $2(1+ q n_q)$. This gives an approximation for \eqref{eq:generalnq} of the following form:
    \begin{align*}
    &\sum_{u\in S_{k-1}(t)}\IP[\text{a friend of individual $u$ is chosen and the edge $u\sim t$ is created}\mid G_t]\\
    &\quad \approx \sum_{u\in S_{k-1}(t)}\sum_{v\in F_{u}(t)}\frac{1}{t}\frac{n_q q}{2(1+ q n_q)} 
    =\sum_{u\in S_{k-1}(t)} \frac{(k-1)}{t} \frac{n_q q}{2(1+ q n_q)} \\
    &\quad=\frac{(k-1)N_{k-1}(t)}{t} \frac{n_q q}{2(1+ q n_q)} .\numberthis\label{eq:secondarynq}
    \end{align*}
The above derivation makes clear why it is possible to produce a full solution to the master equation for $n_q = Deg^R(t)$ (as in  \textcite{lambiotte2016}) and only an approximate solution when $n_q$ is fixed (as in \textcite{Levens2022}). 

The above equations describe how introducing a new individual $t+1$ to an individual of degree $k-1$ leads to an increase in the count of vertices of degree $k$. It is equally important to remember that introducing a new individual $t+1$ to a individual of degree $k$ leads to a corresponding {\it decrease} in the count of vertices of degree $k$. So that, 
when $t+1$ befriends an individual with $k$ friends; given the network at time $t$, the decrease in the expected value of $N_k(t)$ due to that addition is
    \begin{equation}
        \frac{N_k(t)}{t}. \label{eq:deckq}
    \end{equation}
The computation is similar to that of (\ref{eq:secondary}) when individual $t+1$ befriends an individual who is a friend of individual $v$ with degree $k$. For example, in the case where $n_q=Deg^R(t)$, the conditional probability 
for the decrease in expected value of $N_k(t)$ is
    \begin{equation}
        \frac{qkN_{k}(t)}{t}. \label{eq:deckq1}
    \end{equation}

\subsubsection*{Degree of new individuals}
The final thing we need, before we use the equations described here to derive a full master equation for the model, is a description of how the newly added individual changes the degree distribution. 

When $n_q = Deg^R(t)$, individual $t+1$ has degree $k$ if he/she befriends an individual $v$ with $j$ friends, and $k-1$ friends of $v$, but not the remaining $j-k+1$ friends of $v$. For individual $t+1$ to end up befriending $k-1$ friends of $v$, there are $\binom{j}{k-1}$ subsets to choose from. Recalling that the probability that individual $t+1$ connects to $v$ with $j$ friends with probability $N_j(t)/t$, it follows that $t+1$ has $k$ friends (including $v$) with probability
    \begin{equation}
       \frac{N_j(t)}{t} \binom{j}{k-1}q^{k-1}(1-q)^{k-j+1}.
    \end{equation}
Note that when $n_q=1$ then the probability $t$ has one friend is $1-q$ and that he/she has two friends is $q$; and when $n_q\geq 2$, the newly added individual always ends up with $1+n_q$ friends, since $q=1$ in this case. 

\subsubsection*{Combining the terms}

We start with the master equation for the case $n_q = Deg^R(t)$, which can be found in \cite{lambiotte2016} too. Recall that we initiate the network with a triangle, and in view of the network dynamics, there are no isolated individuals. As before, let $G_t$ be the resulting network after $t$ completed steps, and we further let $\IE[N_{k}(t+1)\mid G_{t}]$ be the \emph{expected} value of $N_{k}(t+1)$, \emph{conditional on $G_t$.} Note also the maximum degree of any individual in $G_t$ is at most $t-1$, since an individual can connect to at most $t-1$ individuals in the network. In light of this, we define $N_k(t)=0$ for any $k\geq t$. For any network size/time step $t\geq 3$ and degree $k\geq 1$, we assemble the terms above to obtain
\begin{align*}
    &\IE[N_{k}(t+1)\mid G_{t}]\\
    &=N_{k}(t) + \frac{N_{k-1}(t)(1+q(k-1))}{t} -\frac{N_{k}(t)(1+qk)}{t}\\
    &\quad  + \sum_{j\geq k-1}\binom{j}{k-1} \frac{N_{j}(t)}{t}q^{k-1}(1-q)^{j-k+1}\numberthis \label{eq:specialk=1}\\
    &=N_{k}(t)\bbclc{1+\frac{-(1+qk)+kq^{k-1}(1-q)}{t}} + \frac{N_{k-1}(t)}{t}\clc{(1+q(k-1)) + q^{k-1}}\\
    &\quad + \sum_{j\geq k+1}\binom{j}{k-1} \frac{N_{j}(t)}{t}q^{k-1}(1-q)^{j-k+1}.\numberthis \label{eq:recursionk}
\end{align*}
Define the expected proportion of individuals of degree $k$ at time step $t\geq 3$ as
\begin{equation}\label{eq:exdegprop}
   P_{k}(t)=\frac{\E[N_{k}(t)]}{t}, 
\end{equation}
so that we have the boundary condition
\begin{equation}\label{eq:boundary}
    P_2(3)=1,\quad \text{and}\quad P_j(3)=0,\quad j\ne 2.
\end{equation}
Taking expectation and dividing both sides with $t$ in (\ref{eq:recursionk}), it follows that for $t\geq 3$,
\begin{align*}
    P_k(t+1)&=P_{k}(t)\bbclc{1-\frac{1}{t+1}(2+k(q-q^{k-1}(1-q))}+\frac{P_{k-1}(t)}{t+1}\clc{1+q(k-1)+q^{k-1}}\\
    &\quad +\sum_{j\geq k+1} \binom{j}{k-1}\frac{P_{j}(t)}{t+1}q^{k-1}(1-q)^{j-k+1}.\numberthis\label{eq:mainrecursion}
\end{align*}
for $k\geq1$. This gives a full master equation (without any approximation) for the case $n_q = Deg^R(t)$.

In the case where $n_q$ is fixed we can, as discussed above, only obtain an approximate master equation. Specifically, using \eqref{eq:secondarynq} and the fact that the average degree of the system is $2(1+q n_q)$, we obtain, for $t\geq n_q+2$, 
\begin{align*}
    \E[N_k(t+1)\mid G_{t}] &= N_k(t) + \frac{N_{k-1}(t)}{t} \bbclc{1+\frac{qn_q(k-1)}{2(1+qn_q)}} \\
    &\quad -  \frac{N_{k}(t)}{t} \bbclc{1+\frac{qn_qk}{2(1+qn_q)}},\quad k\geq n_q+1. 
\end{align*}
Recalling that $P_k(t)$ in \eqref{eq:exdegprop} is the expected proportion of individuals of degree $k$ in $G_t$, taking expectation gives
\begin{align*}
   (t+1)P_k(t+1) &= tP_k(t) + P_{k-1}(t) \bbclc{1+\frac{qn_q(k-1)}{2(1+qn_q)}} \\
   &\quad -  P_k(t) \bbclc{1+\frac{qn_qk}{2(1+qn_q)}}, \quad k\geq n_q+1. \numberthis\label{eq:masternq}
\end{align*}
which is equation (3.5) of \cite{Levens2022}, although here we include $n_q$ (and also $q$ in the case of $n_q\geq 2$) and omit the parameter $p$ appearing in \cite{Levens2022}. We do not give the boundary conditions here  because, in what follows we look only at the tail of the distribution where $k\gg n_q$ (but see \cite{Levens2022} for details of the case where $n_q=1$).

\subsubsection{The stationary distribution for $n_q = Deg^R(t)$} 

For the case that $n_q = Deg^R(t)$, we can derive a stationary distribution for $P_k(t)$ as $t \rightarrow \infty$ by closely following the approach in \cite{BarbourL2022}. We denote the expected degree distribution at time step $t\geq 3$ as 
\begin{equation}
    \mathbf{P}(t)^\top:=(P_1(t),P_2(t),\ldots),
\end{equation}
where $P_0(t)$ is removed from above since we know that every individual in the network has degree at least one. Note also $P_k(t)=0$ for all $k\geq t$. 
Then, (\ref{eq:mainrecursion}) can be written in vector notation as
\begin{align}\label{eq:lambtheoreticalsol}
     \mathbf{P}(t+1)^\top= \mathbf{P}(t)^\top\bbclc{I+\frac{[Q]_t}{t+1}}=:  \mathbf{P}(t)^\top A(t), 
\end{align}
where $I$ is the $t\times t$ identity matrix and the matrix $[Q]_t$ is a truncation of the matrix $Q$ with any row $k\geq t+1$ set to be zero. The elements of $Q$ are given by  
\begin{gather*}
    Q_{k,k}=-\clc{2+k(q-q^{k-1}(1-q)},\quad 
    Q_{k,k+1}=1+qk+q^k,\\
    Q_{k,j}=\binom{k}{j-1}q^{j-1}(1-q)^{k-j+1},\quad 1\leq j\leq k-1;\quad  Q_{k,j}=0,\quad j\geq k+2 \numberthis\label{eq:eleQ}
\end{gather*}
for all $k\geq 1$. To solve for $\mathbf{P}(t)^\top$, we can then iterate (\ref{eq:lambtheoreticalsol})  over the time steps with the boundary condition
\begin{equation}
    \mathbf{P}(3)^\top=(0,1,0,\dots).
\end{equation}

We now establish that the expected degree distribution is stationary for all $q\in (0,q^*)$, where $q^*\approx 0.5671$ is the solution to $qe^q=1$. To do so, we first show that given~$q$, there exists~$t_0$ such that for $t\geq  t_0$, $A(t)$ in (\ref{eq:lambtheoreticalsol}) is a stochastic matrix (i.e.\ each of its row sums to one and each element is non-negative). Firstly, summing over all elements of $Q$ in \eqref{eq:eleQ} and then plugging in the sum into $A(t)=I+(t+1)^{-1}[Q]_t$, we get that each row of $A(t)$ indeed sums to one. In view of \eqref{eq:eleQ}, all non-diagonal elements of $A(t)$ are non-negative, and all the diagonal elements of $A(t)$ are non-negative if for all $k\leq t$, 
\begin{equation}\label{eq:pos}
     t-1-kq+kq^{k-1}(1-q)\geq 0; 
\end{equation}
noting that for $k\geq t+1$, the $k$-th diagonal element of $[Q]_t$ is exactly 0. The inequality in \eqref{eq:pos} holds for $k\leq t$ if 
\begin{align}
     t(1-q)-1\geq 0
\end{align}
and this is satisfied if $t\geq (1-q)^{-1}$. Now, given $q<1$, let $t_0$ be a positive integer such that $t_0\geq (1-q)^{-1}+1$, so that $(A(t), t\geq t_0)$ is a family of stochastic matrices, and $\mathbf{P}(t_0)^\top$, the expected degree distribution of the network at time $t_0$, can be obtained from iterating (\ref{eq:lambtheoreticalsol}) with the boundary condition \eqref{eq:boundary}. Therefore, we can consider the \emph{time-inhomogeneous} discrete time Markov chain 
\begin{equation}
    Y=(Y_t,t\geq t_0)
\end{equation}
that
\begin{enumerate}[noitemsep]
    \item lives on the state space $\mathbbm{N}=\clc{1,2,3,\dots}$;
    \item has the initial distribution $\mathbf{P}(t_0)^\top$;
    \item has the transition matrices $(A(t), t\geq t_0)$.
\end{enumerate}
Note that $\mathbbm{N}$ forms a single communicating class for the Markov chain. Since the evolution of the expected degree distribution over time is encoded in \eqref{eq:lambtheoreticalsol}, $Y_t$ has the same probability distribution as the degree of the uniformly chosen individual in the network with $t$ individuals.  

Instead of directly analysing the time-inhomogeneous discrete time Markov chain, we further observe that, $Q$ defined by \eqref{eq:eleQ} is the generator matrix for a homogeneous continuous-time Markov chain (CTMC) on $\mathbbm{N}$. (i.e.\ each of its rows sums to zero and all the non-diagonal elements are non-negative). Hence, let 
\begin{equation}
    X=(X_t,t\geq t_0)
\end{equation}
be a continuous time Markov chain that  
\begin{enumerate}[noitemsep]
    \item lives on the state space $\clc{1,2,3,\dots}$;
    \item has the initial distribution $\mathbf{P}(t_0)^\top$;
    \item has the generator matrix $Q$.
\end{enumerate}

Using the same coupling procedure in \cite[Section 6]{BarbourL2022}, we can construct $X$ and $Y$ such that their trajectories are close enough each other, after rescaling the time of the process $Y$ in a suitable way. The mathematical detail can be found in \cite{BarbourL2022} and is hence omitted here. With this coupling at hand, we can study the process $X$ in place of the original process $Y$. The question on the expected degree distribution is therefore boiled down to a study of a homogeneous continuous time Markov chain. In particular, for the range of $q$ such that the process $X$ is positive recurrent, the degree of the uniformly chosen individual in the network has a stationary distribution. To determine such range of $q$, we can very closely follow the approach of \cite[Theorem 2.1]{BarbourL2022}, which relies heavily on the Foster-Lyapunov-Tweedie techniques (see e.g.\ \cite{menshikov2016} for a book-length treatment on this topic). With a very similar calculation as that for \cite[Theorem 2.1]{BarbourL2022}, we get that the process $X$ is positive recurrent only when  
\begin{equation}
    q<q^*\approx0.5671,
\end{equation}
where $q^*$ is the unique solution to 
\begin{equation}\label{eq:Omega}
    qe^q=1.
\end{equation}
Returning to the problem on the stationarity of the expected degree distribution, the result on the process $X$ (and hence on the process $Y$ through the coupling), the expected degree distribution is therefore stationary {\it only when} $q<q^*\approx0.5671$. In view of the standard properties of a positive recurrent Markov chain, we also emphasise that the stationary distribution does not depend on $\mathbf{P}(t_0)^\top$, and hence the choice of the initial network.

\subsubsection{Power-law distribution for fixed $n_q$}

Assuming that for all values of $0<q\leq 1$, $P_k(t)$ converges to some $P_k:=P_k(\infty)$ as $t\to\infty$ where $(P_1,P_2,\dots)$ is non-defective (i.e.\ $\sum_{k\geq 1}P_k=1$), \cite{Levens2022} used an approximation similar to \eqref{eq:masternq} (but with $n_q=1$) and the mean-field approach to predict a power-law behaviour for the degree distribution. For $n_q\geq 2$ (with $q=1$), we apply the same procedure to \eqref{eq:masternq}. As the argument is entirely similar for $n_q=1$ with $0<q\leq 1$, as before, we leave $q$ in the derivation below to include this case.

Here, we assume that both $t$ and $k$ are continuous and let $dk/dt$ be the increment in degree of an individual of degree $k$.  Writing  \eqref{eq:masternq} as the rate equation for the total rate of increase of degree \(k\)  individuals in one time step, we have:

\begin{equation} \label{rateequation}
    \frac{dk}{dt}=  \frac{1}{t} +  \frac{kqn_q}{2(1+qn_q)t}.
\end{equation}

Using the initial condition that at time step \(t_i\), a new individual with degree $k_i(t_i)$ equalling to the average degree \(2(1+qn_q)\) is added, we solve \eqref{rateequation} to obtain a prediction for the degree $k(t)$ of the individual that is added to the network at time $t$:
\begin{equation} \label{solution1}
    k(t) + \frac{2(1+qn_q)}{qn_q} = \left(2(1+qn_q) +\frac {2(1+qn_q)}{qn_q}\right)\left(\frac{t}{t_i}\right)^\frac{qn_q}{2(1+qn_q)}.
\end{equation}

This solution can also be expressed as
\begin{equation} \label{solution2}
    t_i =\frac{\left(2(1+qn_q) +\frac {2(1+qn_q)}{qn_q}\right)^\frac{2(1+qn_q)}{qn_q}} {\left(k(t) +\frac {2(1+qn_q)}{qn_q}\right)^\frac{2(1+qn_q)}{qn_q}}t.
\end{equation}
Assuming new individuals are added at equal time intervals, the probability density function for any time \(t_i\) within the interval \([t_i, t]\) is

\begin{equation} \label{pdf}
    P(k(t))=\frac{1}{2(1+qn_q) + t}.
\end{equation}
The probability that a node has a degree $k_i(t_i)$ smaller than $k(t)$ is then given by
\begin{equation} \label{pdf1}
    \IP \left[k_i(t_i)<k(t)\right] = 1-\frac{t_i} {2(1+qn_q)+t}.
\end{equation}

Substituting (\ref{solution2}) into (\ref{pdf1}), we obtain
\begin{equation} \label{pdf2}
    \IP \left[k_i(t_i)<k(t)\right] = 1-\frac{\left(2(1+qn_q) +\frac {2(1+qn_q)}{qn_q}\right)^\frac{2(1+qn_q)}{qn_q}t} {(2(1+qn_q)+t){\left(k(t) +\frac {2(1+qn_q)}{qn_q}\right)^\frac{2(1+qn_q)}{qn_q}}}.
\end{equation}

The probability density function  \(P\left( k(t)\right) \) is the partial derivative of equation (\ref{pdf2}) with respect to \(k(t)\). Thus,
\begin{equation} \label{pdf3}
    P\left( k(t)\right) =\frac{\partial \IP \left[k_i(t_i)<k(t)\right]}{\partial k(t)}= \frac{2(1+qn_q)}{qn_q} \frac{\left(2(1+qn_q) +\frac {2(1+qn_q)}{qn_q}\right)^\frac{2(1+qn_q)}{qn_q}t} {(2(1+qn_q)+t){\left(k(t) +\frac {2(1+qn_q)}{qn_q}\right)^{1+\frac{2(1+qn_q)}{qn_q}}}}.
\end{equation}

Assuming that as \(P\left( k(t)\right)\) approaches \(P_k\) ,  \(t \to \infty\), we then have 
\begin{equation}\label{pdf4}
	P_k = \frac{2(1+qn_q)}{qn_q} \left(2(1+qn_q) +\frac {2(1+qn_q)}{qn_q}\right)^\frac{2(1+qn_q)}{qn_q} \left(k +\frac {2(1+qn_q)}{qn_q}\right)^{-\big(1+\frac{2(1+qn_q)}{qn_q}\big)}.
\end{equation} 
In the asymptotic regime of very large $k\gg n_q$, the degree distribution demonstrates a power-law behaviour. Specifically,
\begin{equation}\label{pdf6}
		P_k \sim\left(k +\frac {2(1+qn_q)}{qn_q}\right)^{-\big(1+\frac{2(1+qn_q)}{qn_q}\big)}  \sim k ^{-\big(1+\frac{2(1+qn_q)}{qn_q}\big)} , \quad \text{for all $k$.} 
\end{equation} 
The scaling parameter $\alpha$ can be computed using the exponent of $k$, 
	\begin{equation}\label{alpha}		\alpha = 1+\frac{2(1+qn_q)}{qn_q}=3 +\frac {2}{qn_q}.
\end{equation} 
By adjusting the parameters $q$ and $n_q$, one can therefore obtain a range of `friend of a friend' networks with varying exponents in the interval $3\leq \alpha$.
\subsubsection{Simulations of the degree distribution}
For the cases where $n_q$ is fixed and $n_q=Deg^R(t)$, we now compare the solutions for the expected degree distribution given in \eqref{eq:lambtheoreticalsol} and  \eqref{pdf4} to simulation results\footnote{All the codes for generating the figures in this article can be found in this \href{https://github.com/Watsonlevens/The-properties-of-the-friend-of-a-friend-network-model}{GitHub repository}. }. When $n_q=Deg^R(t)$, with $t$ fixed (but large), we obtain the solution by iterating \eqref{eq:lambtheoreticalsol} over the time steps, with the boundary condition given by \eqref{eq:boundary}. The results for $q=0.3$ and $q=0.8$ are illustrated in panel A and B of Figure \ref{Degree_distribution} below. In Panel A, the observed degree distribution (in blue dotted line) is concentrated around the mean obtained from \eqref{eq:lambtheoreticalsol} (in red solid line); while in Panel B, the observed degree distribution appears to deviate from the mean, especially for smaller degrees. It is possible that the concentration behaviour is only valid for smaller values of $q$, as suggested by the fact that the expected degree distribution is stationary only if $q<q^*\approx0.5671$ and that larger values of $q$ may favour the emergence of `super-hubs' (see e.g.\ \cite{Levens2022} for further empirical studies.) 

When $n_q=1$, a more detailed simulation studies can be found in \cite{Levens2022}. In Panel C below, we focus on $n_q=2$, with $q=1$. The approximation provided in  \eqref{pdf4} appears to capture the shape of the observed line, although for smaller degrees, its slope is slightly steeper than that of the numerical result.


\begin{figure}[htp]
    \centering
    \includegraphics[width=\textwidth]{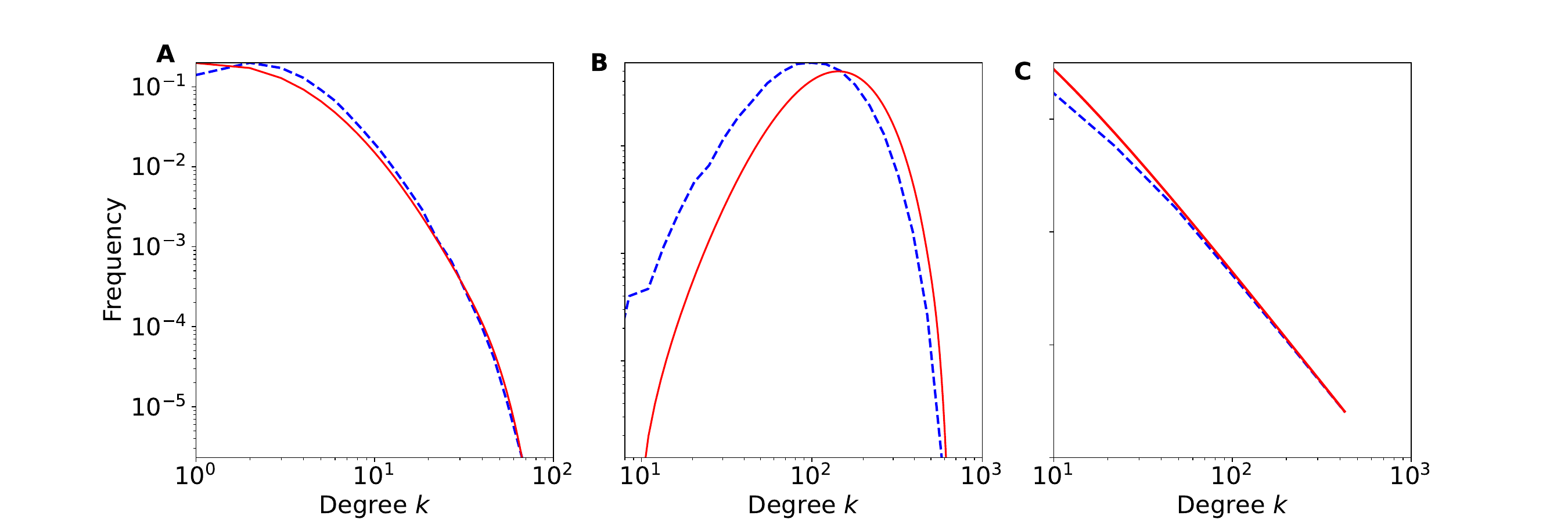}
    \vspace{0.3cm}
    \caption{In Panels A and B, we consider the case where $n_q=Deg^R(t)$, respectively with $q = 0.3$ and $q = 0.8$. In Panel C, we examine the case where $n_q = 2$ and $q = 1$. The red lines in Panels A and B are obtained from iterating \eqref{eq:lambtheoreticalsol} for $t=1000$, with the boundary condition \eqref{eq:boundary}, while the red line in Panel C is given by \eqref{pdf4}. In all three panels, the blue dotted lines represent the simulated degree distributions. Simulations for Panels A and B are run with $t=1000$ over 100 realisations; and for Panel C, it is run with $t=1\times10^6$ over 1 realisation.}
    \label{Degree_distribution}
\end{figure}


\subsection{Triangles}\label{se:tri}

In this section, we investigate the count of triangles, focusing on the case where $n_q\geq 1$ is fixed. We let $\Delta(t)$ be the number of triangles present in $G_t$. At each time step, the network may gain a new triangle from connecting the new individual, the target and one of the $n_q$ friends of the target. As these links are generated independently, for any randomly chosen target, the expected gain in the number of triangles in such a way is distributed as the binomial distribution with parameters $n_q$ and $q$. (When $n_q\geq 2$, $q=1$, so this creates $n_q$ new triangles in the network.) However, when $n_q\geq 2$, the $n_q$ chosen friends of the target can be connected among themselves, and in which case there are additional triangles consisting of two of the $n_q$ friends and the new individual. Let $S_u(t)$ be the set of all possible $n_q$-subsets of the friends of individual $u$ at time $t$, and denote its elements by $\mathbf{i}:=(i_1, i_2,\dots, i_{n_q})$, with $i_1<\dots<i_{n_q}$. Given that individual $u$ is the randomly chosen target of individual $t+1$, the probability that $t+1$ chooses any $\mathbf{i}\in S_u(t)$ is uniform over all possible $\binom{Deg_t(u)}{n_q}$ $n_q$-subsets. The master equation for the case $n_q\geq 2$ is therefore 
\begin{align*}
    \E[\Delta(t+1)\mid G_t] &= \Delta(t) +  n_q \\&\quad + \frac{1}{t}\sum_{u\in V(G_t)}\sum_{\mathbf{i}\in S_u(t)} \sum^{n_q}_{j=1} \sum^{n_q}_{k=j+1}\frac{\mathbbm{1}[\text{$i_j$ and $i_k$ are connected at time $t$}]}{\binom{Deg_t(u)}{n_q}}.\numberthis \label{eq:gentri}
\end{align*}
However, there can be at most $\binom{n_q}{2}$ links among the $n_q$ chosen friends of any randomly chosen target, implying that there can be at most $\binom{n_q}{2}$ additional triangles that do not involve the target individual. Hence, \eqref{eq:gentri} can be bounded as 
\begin{align*}
    \E[\Delta(t+1)\mid G_t] &\leq  \Delta(t) + q n_q + \binom{n_q}{2},
\end{align*}
and iterating the above over the time steps, we obtain the upper bound:
\begin{align*}
    \E[\Delta(t+1)] &\leq  \Delta(n_q+2) + \bigg( n_q + \binom{n_q}{2}\bigg)(t-n_q-2),
\end{align*}
implying in particular the expected number of triangles can at most grow linearly with $t$. We also have the easy lower bound:
\begin{align*}
    \E[\Delta(t+1)\mid G_t] &\geq \Delta(t) +  n_q, 
\end{align*}
which implies 
\begin{align}
    \E[\Delta(t+1)] &\geq \Delta(n_q+2) +  n_q (t-n_q-2).
\end{align}
Combining both bounds,
\begin{align}\label{eq:tribd}
    n_q\leq \liminf_{t\to\infty}\frac{\E\Delta(t)}{t}\leq \limsup_{t\to\infty}\frac{\E\Delta(t)}{t}\leq n_q + \binom{n_q}{2},
\end{align}
implying that $\E\Delta(t)$ increases linearly over time.

\begin{figure}[htp]
    \centering
    \includegraphics[scale=0.3]{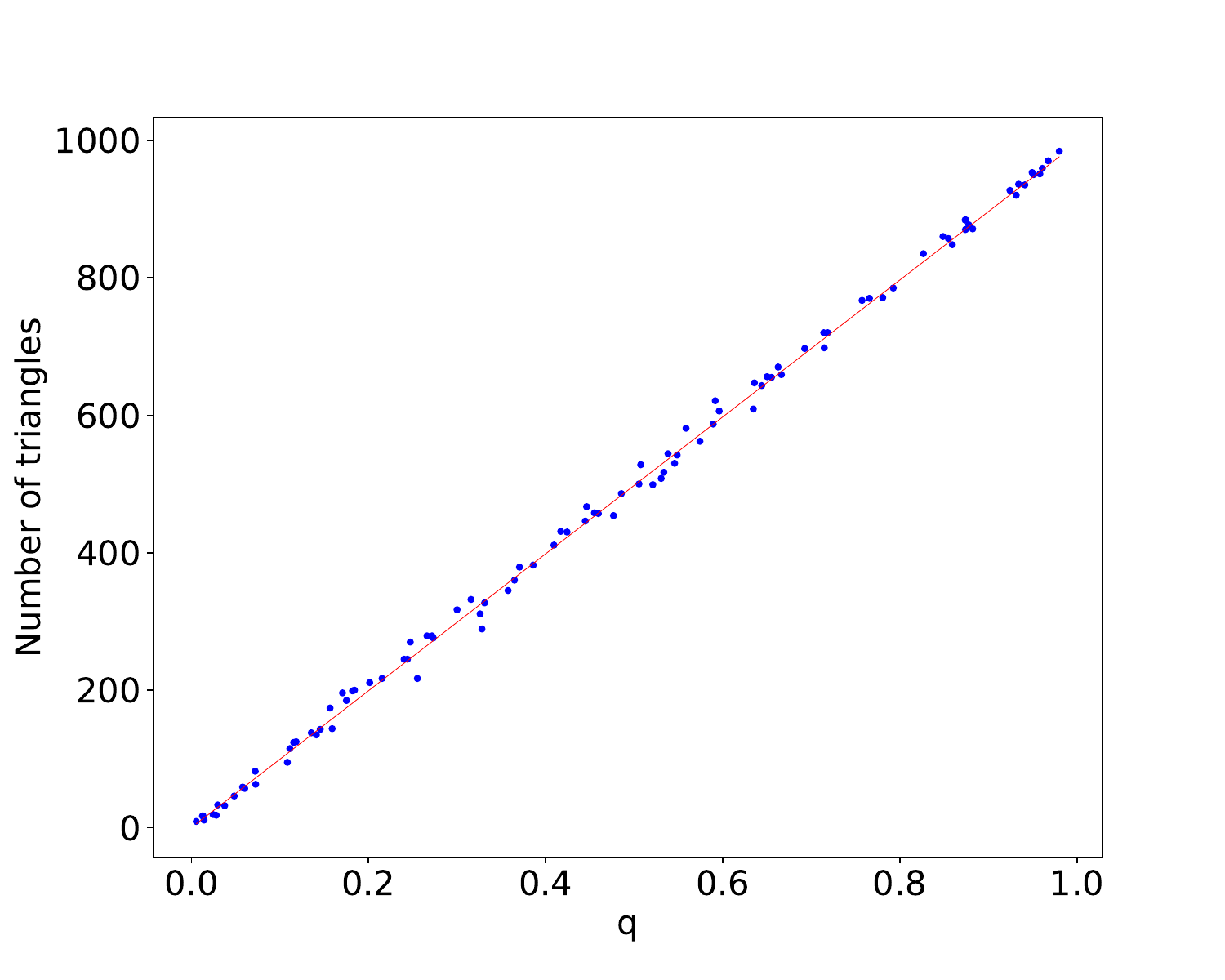}
    \caption{The relationship between the number of triangles and the probability $0<q \leq 1$ for the `friend of a friend' model with $n_q=1$. The red line represents $\E[\Delta(t)]=q(t-3)+1$, the expected number of triangles in the network, and the blue dotted line are simulations run with $t=1000$ and $n_q = 1$.}
    \label{Triangles_distribution}
\end{figure}

When $n_q=1$, the addition of the new individual at each step can lead to at most one new triangle, consisting of the new individual, the target, and the chosen friend of the target. Moreover, observe that the \emph{increment} in the count of triangles at each time step does not depend on the existing graph structure, only on the event that whether the new individual is also connected to the friend of the chosen target (noting that each individual in the network has at least one friend.) More precisely, 
\begin{align}
    (\Delta(t)-\Delta(t-1))_{t\geq 4}\overset{d}{=} (\mathbbm{1}[\text{individual $t$ connects to a friend of its target}])_{t\geq 4}\overset{d}{=} (X_t)_{t\geq 4},
\end{align}
where $\overset{d}{=}$ means equality in distribution, and $(X_t)_{t\geq 4}$ are i.i.d.\ Bernoulli random variables with parameter $q$. Consequently, the total number of triangles at time $t$,
\begin{align} 
    \Delta(t) = \Delta(3)+\sum^t_{j=4}(\Delta(j)-\Delta(j-1))
\end{align}
has the same distribution as $1+Y_t$, where $Y_t$ has the binomial distribution with parameters $t-3$ and $q$. Many asymptotic properties of $\Delta(t)$ follow at once. For instance, the strong law of large numbers would imply 
\begin{equation}\label{eq:triangleconv}
   \lim_{t\to\infty} \frac{\Delta(t)}{t} =q\quad \text{with probability 1,} 
\end{equation}
which is a much stronger result than \eqref{eq:tribd}. Figure \ref{Triangles_distribution} illustrates this linear relationship between the number of triangles and the probability $q$, comparing the expected number of triangles with the corresponding  simulations. Furthermore, it follows from the Central Limit Theorem that as $t\to\infty$,
\begin{align}
    \frac{\Delta(t)-qt}{\sqrt{tq(1-q)}}
\end{align}
converges in distribution to the standard normal distribution. 


The case where $n_q=Deg^R(t)$ has been analysed in \cite{lambiotte2016}, but we reproduce their results here to illustrate how the analysis and the result compare to the case where $n_q$ is fixed. For this version of the model, adding the $(t+1)$-th individual to the network with $t$ individuals, $G_t$, the increase in the number of triangles occurs when
\begin{enumerate}
    \item individual $t+1$ befriends individual $v$ in $G_t$ initially, and a friend $u$ of individual $v$ also becomes friends with individual $t+1$. Let $V(G_t)$ be the set of all individuals in $G_t$ and $F_{v}(t)$ be the set of friends of individual $v$ in $G_t$. Considering all the existing individuals in $G_t$ and all their friends, this happens with the conditional probability given $G_t$
    \begin{equation}
       \sum_{v\in V(G_t)}\sum_{u\in F_{v}(t)} \frac{q}{t} =  \sum_{v\in V(G_t)}\frac{q Deg_t(v)}{t}  = \frac{2q}{t} E(t),
    \end{equation}
    where $E(t)$ is the number of edges in $G_t$ (before adding individual $t+1$.)
    \item individual $t+1$ first befriends an individual $v$ in $G_t$ that belongs to a triangle, and then befriends two other members, say $u_1,u_2$ in the triangle. The edges between individuals $t+1$, $u_1$ and $u_2$ then yield a new triangle. For each of the $\Delta(t)$ triangles in $G_t$, there are three individuals to be chosen as a target. So this occurs with the conditional probability given $G_t$ 
    \begin{equation}
        \frac{3\Delta(t) q^2}{t}.
    \end{equation}
\end{enumerate} 
Hence, the expected number of triangles in $G_{t+1}$, given $G_t$, is given by
\begin{align}
    \E[\Delta(t+1)\mid G_t]
     = \Delta(t) + \frac{3\Delta(t) q^2}{t} + \frac{2q}{t} E(t).
\end{align}
Obtaining the asymptotic behaviour requires us to simultaneously solve the recursion for both $E(t)$ and $\Delta(t)$. As shown in \cite[equation (8)]{lambiotte2016}, this gives
\begin{align}\label{eq:lambtriangle}
    \E[\Delta(t)]\sim
    \begin{cases}
        t\quad &\text{if $0<q<1/2$;}\\
        t\ln t &\text{if $q=1/2$;}\\
        t^{2q}&\text{if $1/2<q<2/3$;}\\
        t^{4/3}\ln t&\text{if $q=2/3$;}\\
        t^{3q^2}&\text{if $2/3<q\leq 1$.}\\
    \end{cases}
\end{align}

\subsection{Clustering}\label{se:cluster}
\subsubsection*{Approximating the clustering coefficient}
In this section, we analyse the cluster behaviour of the `friend of a friend' model when $n_q$ is fixed. We start by defining the clustering coefficient. Given an individual $v$, let $E_v(t)$ be the number of edges that exist between the friends of individual $v$ in $G_t$, the network has $t$ individuals, and $E_v(t)=0$ if $Deg_v(t)=1$. Alternatively, $E_v(t)$ can be thought as the number of triangles that traverse individual $v$ in $G_t$. Recall that $Deg_v(t)$ is the degree of individual $v$ in $G_t$. As observed in the previous section, 
\begin{equation}
    E_v(t) \leq \frac{Deg_v(t)(Deg_v(t)-1)}{2}=\binom{Deg_v(t)}{2},
\end{equation}
where the right-hand side above is the maximum number of edges between the friends of $v$. We define the clustering averaged over all nodes of degree $k\geq 2$ as
\begin{equation}\label{eq:clustering}
    C_k(t) = \frac{2\widetilde E_k(t)}{k(k-1)},
\end{equation}
where 
\begin{equation}
   \widetilde E_k(t) = \frac{1}{N_k(t)} \sum_{v:Deg_v(t)=k} E_v(t)
\end{equation}
is the average number of triangles traversing an individual of degree $k$ in $G_t$. If $k=1$, we set $C_k(t)=0$. 

For the `friend of a friend' model with $n_q=1$ and $q\leq 1$, we derive an exact master equation for 
\begin{align}
    T_k(t):=\sum_{v:Deg_v(t)=k} E_v(t),
\end{align}
and briefly discuss the different approach that was used in \cite{Levens2022}. Observed that $T_k(t)$ increases by $E_v(t)$ if the new individual chooses individual $v$ of degree $k-1$ as its target but does not connect to a randomly chosen friend of individual $v$. The count $T_k(t)$ can increase by  $E_v(t)+1$ too, if the new individual connects to individual $v$ of degree $k-1$ as its target and a randomly chosen friend of individual $v$, in which case there is a new triangle that traverses individual $v$. If one of the $k-1$ friends of $v$ is chosen as a target of $t+1$, and $t+1$ further befriends $v$, in this case $v$ also picks up a new triangle and so this adds $E_v(t)+1$ to $T_k(t)$. On the other hand, $T_k(t)$ decreases by $E_v(t)$, if a individual of degree $k$ is chosen as a target of the new individual, or if a friend of individual $v$ of degree $k$ is chosen as target, and individual $v$ is further chosen as the second friend. Recalling that $Ne_v(t)$ is the set of neighbours of $v$ in $G_t$, the resulting master equation is therefore
\begin{align*}
    \E[T_k(t+1)\mid G_{t}] &= T_k(t) + \sum_{v:Deg_v(t)=k-1}\frac{1}{t}((1-q)E_v(t) + q(E_v(t)+1))\\
    &\quad + \sum_{v: Deg_v(t)=k-1} \sum_{u\in Ne_v(t)} \frac{(E_v(t)+1)}{t}\frac{q}{Deg_u(t)} \\
    &\quad -\sum_{v:Deg_v(t)=k}\frac{E_v(t)}{t} - \sum_{v: Deg_v(t)=k} \sum_{u\in Ne_v(t)} \frac{E_v(t)}{t}\frac{q}{Deg_u(t)}\\
    &=T_k(t)\bbclr{1-\frac{1}{t}}+\frac{1}{t}T_{k-1}(t) +qP_{k-1}(t)\\
    &\quad + \sum_{v: Deg_v(t)=k-1} \sum_{u\in Ne_v(t)} \frac{(E_v(t)+1)}{t}\frac{q}{Deg_u(t)} \\
    &\quad  - \sum_{v: Deg_v(t)=k} \sum_{u\in Ne_v(t)} \frac{E_v(t)}{t}\frac{q}{Deg_u(t)}.\numberthis\label{eq:triangledegk}
\end{align*}
Similar to the master equation for the expected degree distribution, the dependence on $Deg_u(t)$ and $P_k(t)$ make the analysis of the master equation above difficult. If we proceed as in the case of the expected degree distribution, approximating $Deg_u(t)$ with the average degree $2(1+q)$, then with a further calculation, the approximation takes the form
\begin{align}
     \E[T_k(t+1)\mid G_{t}]&\approx T_k(t) + \frac{1}{t}(T_{k-1}(t)-T_k(t)) \nonumber \\
     &\quad + \frac{q}{2(1+q)t}((k-1)T_{k-1}(t)-kT_k(t)) + \bbclr{q+\frac{q(k-1)}{2(1+q)}}P_{k-1}(t)
     ,
\end{align}
which still involves the term $P_{k-1}(t)$. We thus provide a more rigorous derivation of the master equation for the triangle count given in \textcite{Levens2022}.


When $q=1$ and $n_q=1$, \cite{Levens2022} has argued that $C_k(t)$ is \emph{approximately} $2/k$ but their simulation result suggests that it is in fact exact. Here we provide the additional mathematical details behind this. More specifically, we prove that for all $t\geq 3$,
\begin{align}\label{eq:indhyp}
    E_v(t) = Deg_v(t)-1\quad\text{for all $v\in\{1,\ldots,t\}$},
\end{align}
which in turn implies that
\begin{align}
    \widetilde E_k(t) = \frac{1}{N_k(t)} \sum_{v:Deg_v(t)=k} (Deg_v(t)-1) = k-1
\end{align}
so by \eqref{eq:clustering},
\begin{equation}
    C_k(t) = \frac{2}{k}, 
\end{equation}
as required.

We prove \eqref{eq:indhyp} by an induction over the network size $t$. For the base case of the induction ($t=3$,) recall that we initiate the network with a triangle when $n_q=1$, so \eqref{eq:indhyp} follows immediately. Now, assume that the induction hypothesis \eqref{eq:indhyp} is valid for some $t>3$. When constructing the network with $t+1$ individuals, we connect the new individual to the randomly chosen target, say~$v$, it follows that (after completion of this step)
\begin{align}
    Deg_v(t+1) = Deg_v(t) + 1.
\end{align}
Furthermore, the new individual $t+1$ connects to a randomly chosen friend, say $u$, of~$v$ with probability one. So using \eqref{eq:indhyp}, 
\begin{align}
    E_v(t+1) = E_v(t) + 1 = Deg_v(t)-1 + 1 = Deg_v(t) = Deg_v(t+1)-1.
\end{align}
As for individual $u$, by the same consideration as above,
\begin{align}
    Deg_u(t+1) = Deg_u(t) + 1\quad \text{and}\quad  E_u(t+1) =Deg_u(t+1)-1.
\end{align}
For individual $t+1$, we observe that with probability one,
\begin{equation}
    Deg_{t+1}(t+1)=2,
\end{equation}
and because $u$ and $v$ are connected, 
\begin{equation}
    E_{t+1}(t+1)=1,
\end{equation}
thus proving \eqref{eq:indhyp} for the new individual $t+1$. As for individual $w$ that is not $u,v$ and $t+1$, we have $E_w(t+1)=Deg_w(t+1)-1=Deg_w(t)-1=E_w(t)$ (by the induction hypothesis), since they are not part of the newly formed triangle. This completes the induction step and proves \eqref{eq:indhyp}. 




When $n_q\geq 2$ (and $q=1$), it is much harder to rigorously analyse the clustering behaviour of the model. In this direction, we instead consider the clustering behaviour of the randomly chosen target at each step. To do so, we first consider a modification of the model with the same parameters $n_q\geq 2$ and $q=1$. To slightly simplify our argument below, we also start with a fully connected graph of $t_0=n_q+1$  individuals (instead of $n_q+2$ individuals). In the modified version, one of the individuals in the initial network, say individual 1, is always chosen as the target at each time step, so that the newly added individual always connects to the person and to a random $n_q$-subset of his/her friends. Saving notation, let $E_1(t)$ be the number of triangles at time $t$ that contains individual 1 and let $Deg_1(t)$ be the degree of individual 1 at time $t$ in this modified network. We first examine how the clustering measure of individual 1, namely 
\begin{equation}\label{eq:cc}
    \widehat C_1(t)=\begin{cases}
         \frac{2E_1(t)}{Deg_1(t)(Deg_1(t)-1)} &\text{if $Deg_1(t)\geq 2$,}\\
         0&\text{if $Deg_1(t)\leq 1$.}
    \end{cases}
\end{equation}
evolves over time (or equivalently, as its degree increases). In the model above, $Deg_1(t)\geq n_q$ for all $t\geq t_0$. Counting all pairs of individuals in the complete graph, excluding individual 1, we have 
\begin{equation}
    E_1(t_0) = \binom{n_q}{2}
\end{equation}
and clearly
\begin{equation}
     Deg_1(t_0)=n_q.
\end{equation}
For $t=t_0+i$, the degree of individual 1 increases by one when we add individual $t+1$ to the system of $t$ individuals, due to the new edge linking him/her and $t+1$. On the other hand, the number of triangles containing individual 1 increases by $n_q$, since $t+1$ is now a friend of individual 1, and with probability $q=1$, it connects to a random $n_q$-subset of friends of individual 1 in the existing graph. Hence, 
\begin{equation}
    E_1(t+1) = E_1(t) + n_q,\quad Deg_1(t+1)=Deg_1(t) + 1.
\end{equation}
Iterating, we conclude that 
\begin{equation}\label{eq:trideg}
    E_1(t) = E_1(t_0) + n_q\cdot i, \quad Deg_1(t)=Deg_1(t_0) + i,
\end{equation}
which also implies that 
\begin{equation}\label{eq:predtriangles}
     E_1(t) = \binom{n_q}{2} +n_q (Deg_1(t)-n_q).
\end{equation}
Applying \eqref{eq:trideg} and $t=t_0+i$ to the clustering measure of individual 1 in \eqref{eq:cc}, we obtain
\begin{align} 
    \widehat C_1(t)= \frac{2\{E_1(t_0) + n_q\cdot (t-t_0)\}}{(Deg_1(t_0)+t-t_0)(Deg_1(t_0)+t-t_0-1)}\sim \frac{2n_q}{t}.
\end{align}

We now use \eqref{eq:trideg} to predict the number of triangles that traverse the target individual at each step of the \emph{original} `friend of a friend' model, where the target is resampled at each time step. A direct analysis of this model faces an additional complication, arising from the fact that an individual may gain new friends and triangles whenever his/her friend are chosen as targets, before or after him/herself is selected. Therefore, denote $E^R(t)$ the number of triangles that traverses the randomly chosen individual $R=R(t)$ at time $t$ and recall that $Deg^R(t)$ is the degree of individual $R$, we use the following approximation given by \eqref{eq:predtriangles}:
\begin{equation}\label{eq:pt1}
     E^R(t) \approx \binom{n_q}{2} +n_q (Deg^R(t)-n_q).
\end{equation}
\begin{figure}[htp]
    \centering
    \includegraphics[scale=0.8]{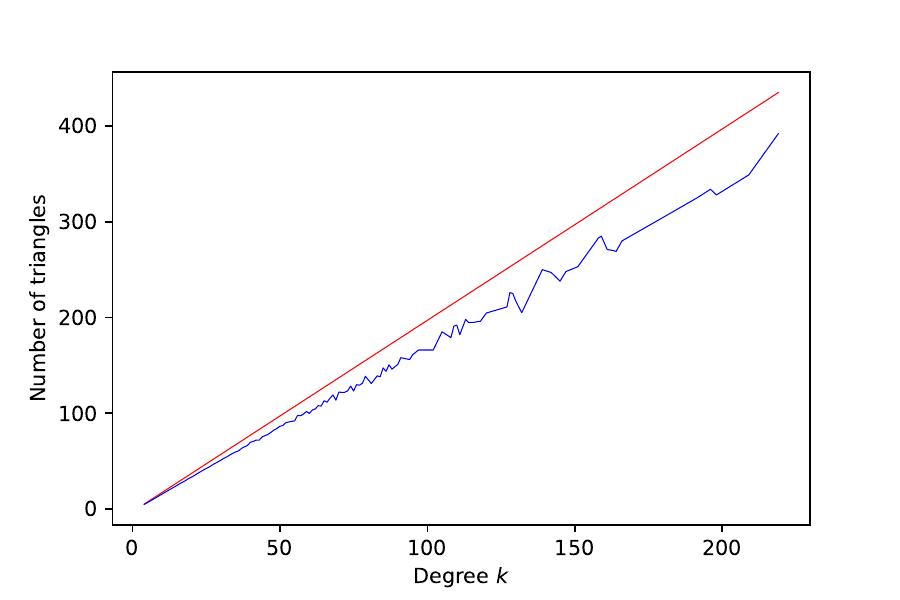}
    \vspace{0.3cm}
    \caption{The average number of triangles for targets with degree k (\ref{eq:pt1}) for the `friend of a friend' model generated with the parameters $q=1$ , $n_q=2$ and $N=100000$. The red line shows the predicted number of triangles for a target that has degree k when it is  chosen, and the blue line shows the observed numbers of triangles passing through the targets of degree $k$, averaged over the number of times we choose a target of degree $k$.}
    \label{Worsttriangles_distribution}
\end{figure}


In Figure \ref{Worsttriangles_distribution} above, we compare the predicted and the actual numbers of triangles passing through the randomly chosen target individual at each time step when growing the original `friend of a friend model' with $n_q=2$ and $q=1$. The red line is the number of triangles predicted by \eqref{eq:pt1} and the blue line is the observed number of triangles, which for each degree $k$, is obtained from averaging the numbers of triangles traversing the targets of degree $k$ over the number of steps at which we choose a target of degree $k$. There appears to be an overestimation of the triangle count that worsens as the degree increases, but is otherwise good for small values of $k$. This is perhaps due to the fact that targets with smaller degrees are less likely to gain triangles before by connecting to individuals who choose their friends as targets, so the numbers of triangles traversing these small degree targets should be closer to the corresponding predicted values. However, the approximation in \eqref{eq:pt1} seems to capture the trend well. The fluctuation in the triangle count also seems to be rather small. Summarising, the result suggests that at least for $n_q=2$, the clustering of the targets are somewhat robust to the additional randomness from choosing the target at each step uniformly from the existing individuals.

\section{Discussion}\label{se:discussion}

We have studied the intuitively appealing `friend of a friend' model for social networks, where the newly added individual not only attaches to the randomly chosen target in the existing network, but also to the $n_q$ chosen friends of the selected individual, each with probability $0<q\leq 1$. The parameter $n_q$, can either be fixed or equal to the degree of the chosen target at the time, and so the model encompasses the ones studied in \textcite{Levens2022} and \textcite{lambiotte2016}. As these are models for understanding the emergence of real-world networks, which are commonly characterised by power-law degree distribution and small-world property, we focus on studying the degree distribution and the clustering behaviour of the model, therefore extending the results of \cite{lambiotte2016,Levens2022}. Furthermore, we also investigate the triangle count of the model, which is also useful for network comparisons and understanding the structural properties of the model itself.

As pointed out in the introduction, the newly added individuals in the model may not have enough individuals to connect to when $n_q\geq 2$ and $q<1$. It is possible to remedy this by further specifying the network growth rule when the target has less than $n_q$ friends to choose from. For instance, at each step $i\leq n_q$, the new individual may simply sample a friend of the target with replacement, so that he/she may connect to the same individual more than once (with the multiple-edges merged.) We leave these and other modifications of the model for future research. 

Below we discuss in more detail the connections of our results to the existing ones.

\subsubsection*{The degree distributions}

In the analysis of the master equation of the expected degree distribution, we emphasise that when $n_q$ is equal to the degree of the randomly chosen target individual, no approximations and assumptions are made in our analysis. This is in contrast to \cite{lambiotte2016}, where it is assumed that $P_k(t)=N_k(t)/t$ do not depend on $t$ when the network is sparse and large; and an approximation for the term
\begin{equation}
    \sum_{j\geq k+1}\binom{j}{k-1} \frac{N_{j}(t-1)}{t-1}q^{k-1}(1-q)^{j-k+1}
\end{equation}
is applied. It is also predicted in \cite{lambiotte2016} that the expected degree distribution has a power-law behaviour when $0< q<1/2$, where the exponent satisfies 
\begin{equation}
    \gamma = 1+q^{-1}-q^{\gamma-2}
\end{equation}
with one of the solutions being $\gamma>2$. We believe that this predicted power-law behaviour in fact holds for $0< q<q^*\approx 0.5671$, and can be established rigorously using the argument of \cite{jacquet2020} that is applied to a very similar model, where the only difference is that, at each step, the new individual does not connect to the randomly chosen target. 


The property of the degree distribution of the model when $n_q$ is fixed, has been studied by comparing the simulation result to the power-law behaviour predicted by
 the mean-field approximation (\ref{pdf4}). The comparison suggests that the prediction indeed captures the shape of the degree distribution, especially when the degree is large. We also note that while the mean degree of the network changes as $n_q$ changes, the way in which the power-law exponent in (\ref{alpha}) depends on $n_q$ is still similar to that in \cite{Levens2022}.  
\subsubsection*{Triangle count}
By deriving the master equation for the triangle count in the case where $n_q$ is fixed, we show that the triangle count of the model has a very different growth behaviour when compared to the case where $n_q$ is equal to the degree of the randomly chosen target at each step. When $n_q=1$, we further prove a Central Limit Theorem and a strong law of large number for the triangle count, which can be used for building a statistical test for assessing the suitability of the model in fitting an observed network. As discussed in the derivation of the master equation \eqref{eq:gentri}, it is much harder to obtain finer results such as the asymptotic distribution when $n_q\geq 2$, and we leave them for future research.

\subsubsection*{Clustering}
We also investigate the clustering behaviour of the model when $n_q$ is fixed, considering in particular how the clustering of the individuals changes with respect to their degrees. In particular, we use a suitable master equation to illustrate the general difficulty in the analysis of clustering. We also show that a precise result can be obtained when $n_q=1$ with $q=1$; and when $n_q\geq 2$ with $q=1$, a reasonable approximation of the clustering of the randomly chosen targets can be achieved. It would be interesting to further investigate the clustering behaviour of the model. For example, one might consider a more detailed analysis of the clustering of the randomly chosen targets, an analytical investigation of how fixing different values of $n_q$ impacts the total clustering coefficient of the model, as well as the clustering behaviour of the model when $n_q$ is the same as the degree of the random target at each step.



\paragraph{Acknowledgement.} TYYL is supported by the Knut and Alice Wallenberg Foundation, 
 the Ragnar S\"oderberg Foundation and 
the Swedish Research Council.

\nocite{*}
\printbibliography[heading=bibintoc]

\end{document}